\documentclass[10pt,letterpaper]{article}
\usepackage{opex3}
\usepackage[draft]{hyperref}
\usepackage{amsmath}
\usepackage{ae} 

\begin{document}

\title{High-bandwidth squeezed light at 1550\,nm from a compact monolithic PPKTP cavity}

\author{Stefan Ast,$^1$ Moritz Mehmet,$^{1,2}$ and Roman Schnabel$^{1,*}$}

\address{$^1$Max Planck Institute for Gravitational Physics, Albert Einstein Institute, and Institut f\"{u}r Gravitationsphysik,
Leibniz Universit\"{a}t Hannover, Callinstrasse 38, D-30167 Hannover, Germany
\\
$^2$Centre for Quantum Engineering and Space-Time Research - QUEST, Leibniz Universit\"{a}t Hannover, Welfengarten 1, 30167 Hannover, Germany
}

\email{$^*$roman.schnabel@aei.mpg.de} 



\begin{abstract}
We report the generation of squeezed vacuum states of light at 1550\,nm with a broadband quantum noise reduction of up to 4.8\,dB ranging from 5\,MHz to 1.2\,GHz sideband frequency. We used a custom-designed 2.6\,mm long biconvex periodically-poled potassium titanyl phosphate (PPKTP) crystal. It featured reflectively coated end surfaces, 2.26\,GHz of linewidth and generated the squeezing via optical parametric amplification. Two homodyne detectors with different quantum efficiencies and bandwidths were used to characterize the non-classical noise suppression. We measured squeezing values of up to 4.8\,dB from 5 to 100\,MHz and up to 3\,dB from 100\,MHz to 1.2\,GHz. The squeezed vacuum measurements were limited by detection loss. We propose an improved detection scheme to measure up to 10\,dB squeezing over 1\,GHz. Our results of GHz bandwidth squeezed light generation provide new prospects for high-speed quantum key distribution.
\end{abstract}




\section{Introduction}

Quantum key distribution (QKD) offers the possibility of unconditionally secure data transmission. Squeezed states of light can be employed as a resource for entanglement-based QKD in the continuous variable (CV) regime \cite{PhysRevA.61.010303, Madsen2012, Hillery2000, Silberhorn2002}.

The effective data rate for entanglement-based QKD depends on both, the squeezing strength and the squeezing bandwidth. A higher squeezing strength increases the average number of bits per measurement \cite{Cerf2005}. The measurement rate (speed) increases linearly with squeezing bandwidth. High squeezing values as well as a high bandwidth are therefore both vital for high-speed QKD based on non-classical states.

Standard telecom fibers offer the possibility to distribute the quantum states \cite{PhysRevA.76.042305,Momo2010} for QKD and are already available for communication networks. Their low loss at 1550\,nm laser wavelength is advantageous to protect the state from decoherence \cite{Tobi2011, PhysRevLett.102.130501}, degradation of security and reduction of data rate. But the decoherence in a km scale fiber network will still be a limiting factor. Since squeezing strength is highly degraded by optical loss in these fibers, the QKD data rate will not be increased further by pushing the squeezing value to more than 10\,dB. The squeezing bandwidth is not significantly influenced by optical loss in fibers and enables an increase in data rate via high measuring speeds. Therefore, a high squeezing bandwidth can compensate the limitations in squeezing strength for real QKD applications.

Slusher \emph{et al.} demonstrated the generation of squeezed states for the first time in 1985 \cite{Slu85}. Continuous-wave squeezed vacuum states are most successfully produced via parametric down-conversion in second-order nonlinear crystals \cite{PhysRevA.29.408,PhysRevLett.57.2520}. The crystal is placed inside an optical resonator to enhance the parametric process. Based on such a setting, a non-classical noise suppression of more than 12.3\,dB \cite{Momo2012-12dB-1550} at 1550\,nm over several tens of MHz was recently achieved. The bandwidth, however, is decreased by an increasing resonator enhancement.

A squeezed vacuum state with more than 2\,GHz, but merely 0.3\,dB of non-classical noise suppression was recently measured \cite{Ast-Hem-GHz-Sqz}. The experiment used no enhancement resonator for the fundamental squeezing field. This allowed the very high bandwidth, sacrificing high squeezing factors at the same time.

The combination of both, a high squeezing factor as well as a high bandwidth in the GHz range, can be achieved using a trade-off between high cavity linewidth and resonant enhancement. A non-classical noise suppression of up to 11.5\,dB at 1064\,nm over 100\,MHz was measured in \cite{PhysRevA.81.013814}.

In the experiment presented here, we used a 2.6\,mm long monolithic crystal cavity with a low finesse of about 14 for the fundamental wavelength. It was custom-designed for the purpose of generating squeezed light with more than 2\,GHz bandwidth as well as a high squeezing factor at 1550\,nm. We observed a non-classical noise suppression of up to 4.8\,dB from 5 to 100\,MHz and up to 3\,dB from 100\,MHz to 1.2\,GHz. Different from \cite{Senior2007}, we observe \emph{broadband} GHz squeezing from a single longitudinal mode of the squeezing resonator. The measured squeezing was limited mainly by the homodyne detection efficiency and detector dark noise, while the squeezing resonator produces significantly stronger squeezing over the whole linewidth of 2.26\,GHz.

\section{Experimental setup}
\label{sec:exp}
\begin{figure}[bt]
\centerline{\includegraphics[width=\textwidth]{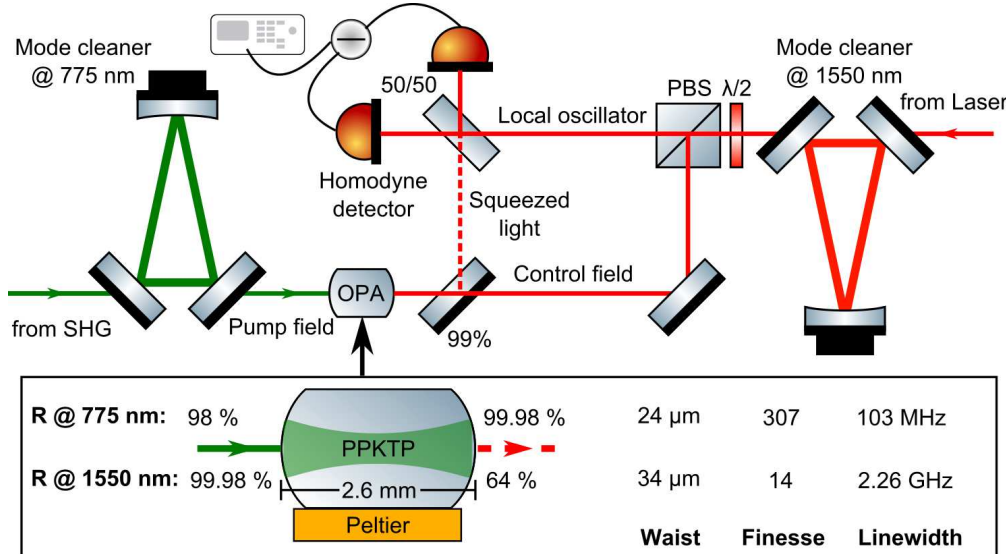}}
\caption{Schematic of the experimental setup. Mode cleaner: three-mirror ring cavity for spatial mode filtering; OPA: monolithic PPKTP cavity (squeezing resonator) with reflective coatings on crystal surfaces; control field: reference beam for alignment of the squeezed vacuum mode onto the balanced homodyne detector; PBS: polarizing beam splitter.}
\label{fig:Mono-GHz-Sqz-Setup}
\end{figure}

A schematical overview of our experimental setup is shown in Fig. \ref{fig:Mono-GHz-Sqz-Setup}. The laser source was an erbium-doped fiber laser with an output power of 1.3\,W at 1550\,nm. A three-mirror ring cavity served as a spatial mode cleaning device and supressed amplitude- and phase noise above its linewidth of about 2.6\,MHz. A fraction of the power was transmitted towards the balanced homodyne detector as a local oscillator beam as well as a control beam for the alignment of the squeezed field. 

The remaining 1.1\,W were converted into about 1\,W at 775\,nm inside a second-harmonic generation resonator (SHG). Details on the SHG are described in \cite{Momo2012-12dB-1550}. A second three-mirror ring cavity filtered the light at 775\,nm, which was then coupled into the optical parametric amplification (OPA) squeezing resonator. 

The OPA consisted of a biconvex monolithic PPKTP crystal with reflective coatings for 775/1550\,nm and radii of curvature of 12\,mm on its surfaces. The crystal coating's reflectivities were $\text{R}_{1}= 99.98\,\%$, $\text{R}_{2}= 64\,\%$ at 1550\,nm and $\text{R}_{1}= 98\,\%$, $\text{R}_{2}= 99.98\,\%$ at 775\,nm, respectively. Together with the short crystal (cavity) length of 2.6\,mm, this led to a high resonator bandwidth of about 2.26\,GHz, a finesse of about 14 and a free spectral range of about 31.75\,GHz. 

The finesse measured for the 775\,nm pump field was 307, corresponding to a resonant enhancement of the intra-cavity power by a factor of about 100. The squeezing resonator was pumped with an internal power of up to 37\,W (375\,mW external). This was below the OPA threshold of 65\,W, which was numerically simulated taking into account the intra-cavity waist sizes of $\text{w}_{1550}=\rm 33.86\,\mu\text{m}$ and $\text{w}_{775}=\rm 23.94\,\mu\text{m}$, the crystal's effective nonlinearity of $\text{d}_{\rm eff}=\rm 7.3\,\text{pm}/\text{V}$ and the crystal`s absorption for both wavelengths. We measured the absorption for PPKTP to be $\alpha_{1550}=84\pm 40\,\text{ppm}/\text{cm}$ and $\alpha_{775}=125\pm26\,\text{ppm}/\text{cm}$ for the fundamental and harmonic wavelengths as reported in \cite{J-Steinlechner-Abs-2013}. 

We observed a small thermally induced deformation of the cavity mode and second-harmonic generation to 387.5\,nm in the squeezing resonator. To avoid damaging the crystal, we did not increase the pump power to more than 375\,mW.

A Peltier-element stabilized the crystal temperature to cavity resonance. A feed-back control loop was not necessary due to the intrinsically stable, short monolithic cavity design. At the temperature chosen, the crystal also was close to its quasi-phase matching temperature of about $46\,^{\circ}\text{C}$.

The squeezed vacuum state left the crystal on the opposite side through which the pump field entered. It propagated through several mode-matching optics and was detected with a balanced homodyne detector. A control field at 1550\,nm served as a spatial reference for the propagating squeezed vacuum and was used to adjust the homodyne detector visibility. We blocked the control field while performing the squeezed-light measurement.

The propagation loss was estimated to be about $\text{L}=\rm 10\,\%$ due to non-perfect HR/AR coatings on mirrors and lenses. The homodyne visibility was measured to be $\beta=90\,\%$ and was mainly limited by mode-mismatch induced higher order modes in the control beam that was reflected off the squeezing resonator. Without taking the photo diode quantum efficiency into account, the detection efficiency was $\eta=\beta^{2}\cdot (1-\text{L})= 0.73$. The outcoupling effciency of our squeezing resonator was close to unity and was therefore negligible in our experiment. Two different homodyne detectors were used to perform two subsequent squeezed vacuum measurements at different sideband frequencies.

\section{GHz-bandwidth measurement}
\label{GHz-freq}
\begin{figure}[!htbp]
\centerline{\includegraphics[width=\textwidth]{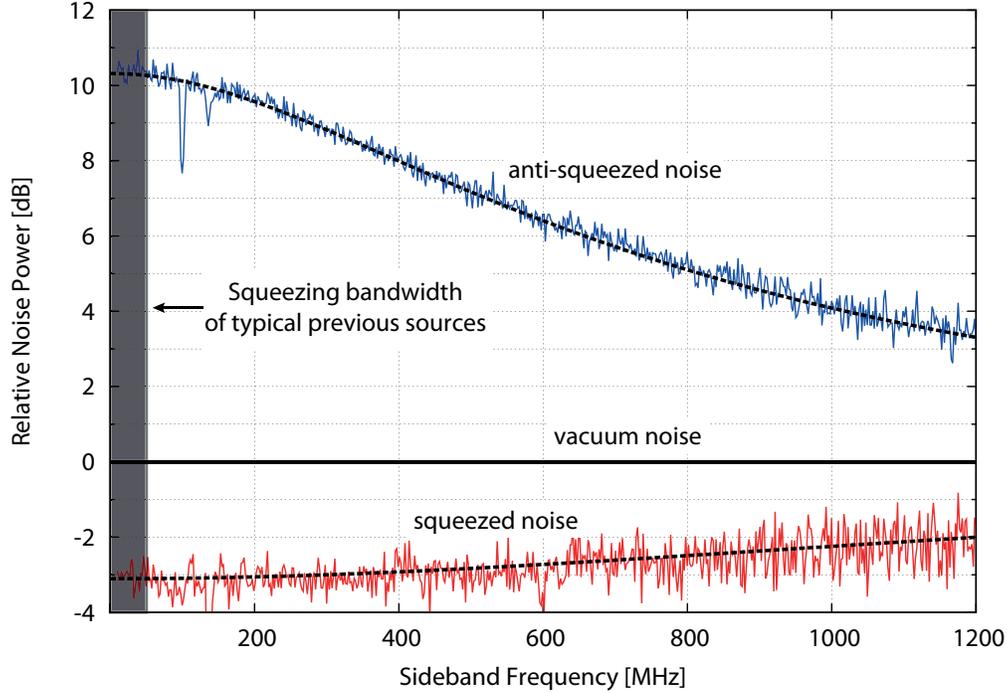}}
\caption{GHz bandwidth measurement. Squeezed-vacuum measurement from 10\,MHz to 1.2\,GHz sideband frequency using the balanced photo receiver \emph{New Focus} type \emph{1617-AC FS}. We measured a squeezing (red) of up to 3\,dB as well as anti-squeezing (blue) of up to 10.4\,dB  above vacuum noise (black). Squeezing decreased to 2\,dB and anti-squeezing to 3.5\,dB due to the finite cavity linewidth of the squeezing resonator. The measurements shown are dark-noise corrected.  The dark-noise clearance was merely 7\,dB at 5\,MHz decreasing to 3\,dB at 1\,GHz. A typical squeezed light source as in \cite{Momo2012-12dB-1550,Senior2007,McKenzie2004,Breitenbach1995} yields a several ten-times smaller squeezing bandwidth, which is highlighted here in grey. The total detection efficiency of our system was fitted to be 53\,\%. A numerical simulation used all given parameters to fit the measured squeezing (dashed black).}
\label{fig:High-frequency-Sqz-same-arm}
\end{figure}
The GHz-bandwidth measurement was performed using a balanced photo receiver from \emph{New Focus}, type \emph{1617-AC FS}. We performed a squeezed-vacuum measurement using 2\,mW local oscillator, 375\,mW harmonic pump power for the squeezing resonator, a resolution bandwidth (RBW) of 5\,MHz, a video bandwidth (VBW) of 3\,kHz and a sweep time of 170\,ms. The dark-noise clearance with 2\,mW local oscillator power ranged from 7\,dB at 10\,MHz to around 2.5\,dB at 1.2\,GHz.

The directly observed squeezing was about 2.5\,dB at 10\,MHz and about 1\,dB at 1.2\,GHz limited partly by the low dark noise clearance of the balanced photo detector. Figure~\ref{fig:High-frequency-Sqz-same-arm} shows a dark-noise corrected squeezing spectrum from 10\,MHz to 1.2\,GHz normalized to the vacuum noise level. We measured a squeezing value of up to 3.0\,dB and an anti-squeezing value of up to 10.4\,dB. The squeezed vacuum decreased to 2\,dB and the anti-squeezed vacuum decreased to 3.5\,dB at sideband frequencies of 1.2\,GHz due to the finite linewidth of 2.26\,GHz for the squeezing resonator.

The measured values correspond to a total detection efficiency of about 0.53. Using the homodyne efficiency of  $\eta=\beta^{2}\cdot (1-\text{L})= 0.73$, we deduced a photo diode detection efficiency of again 73\,\%. The dark-noise corrected squeezing strengths are thus mainly limited by optical losses in the detection process.

We numerically simulated the measurements with the given parameters for the cavity, the pump field and the homodyne detection efficiency using the nonlinear cavity simulator (N.L.C.S.) by Nico Lastzka. The simulations are in very good agreement with the measured spectra. They include the decreasing noise suppression due to the cavity linewidth limitation and can be found as dashed lines in Fig. \ref{fig:High-frequency-Sqz-same-arm}.

\section{100 MHz bandwidth measurement}
\label{sec:MHz-freq}
\begin{figure}[!htbp]
\centerline{\includegraphics[width=\textwidth]{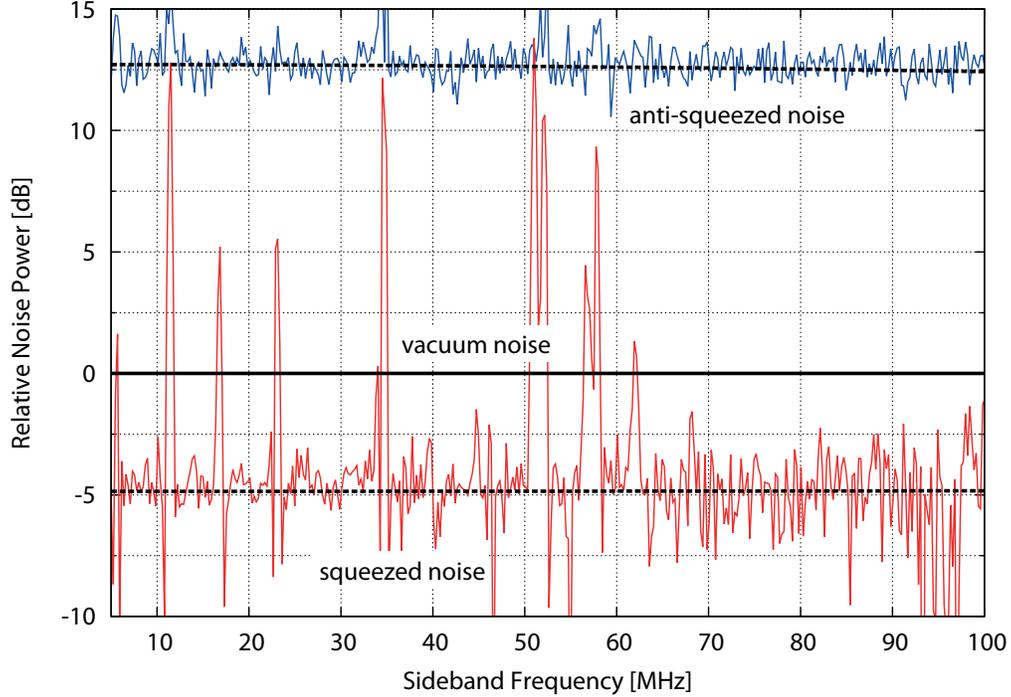}}
\caption{MHz bandwidth measurement. Squeezed-vacuum measurement from 5 to 100\,MHz sideband frequency using a homodyne detector with 99\,\% quantum efficiency. We measured squeezing (red) of 4.8\,dB and anti-squeezing (blue) of 12.7\,dB  with respect to the vacuum noise level (black). The measurement is dark-noise corrected. The measured squeezing below 20\,MHz is, however, not influenced by the dark noise correction due to the detector's low dark noise at low frequencies. The total detection efficiency was fitted to be 72.5\,\%. The dashed black lines correspond to our numerical simulation. The peaks in the squeezing spectrum originated from electronic pick-up of the homodyne detector due to antenna effects and are also visible in the detector`s dark noise.}
\label{fig:Low-frequency-measurement}
\end{figure}
Our second measurement used a home-made homodyne detector based on custom-made high quantum efficiency photo diodes and covered sideband frequencies between 5 and 100\,MHz. The photo currents were directly subtracted, electronically amplified and fed forward to the spectrum analyser. This allowed local oscillator powers as high as 5\,mW saturating neither the homodyne detector's photo diodes nor the operational amplifiers. Due to the high electronic gain and the high local oscillator power, the detector yielded a dark-noise clearance of 20\,dB at 5\,MHz decreasing to less than 3\,dB at 100\,MHz. The high transimpedance gain as well as the finite speed of the photo diodes were responsible for the decreasing dark-noise clearance and limited the detector's bandwidth. At frequencies above 100\,MHz the low dark-noise clearance prevented useful squeezing measurements.

Our squeezed-vacuum measurement used a 5\,mW local oscillator power, 375\,mW harmonic pump power for the squeezing resonator, a RBW of 500\,kHz, a VBW of 3\,kHz and a sweep time of 170\,ms. The measurements without dark-noise correction yielded a non-classical noise suppression of about 4.8\,dB at 5\,MHz and about 2\,dB at 100\,MHz, which was limited by a dark-noise clearance as low as 3\,dB. Figure \ref{fig:Low-frequency-measurement} shows dark noise corrected measurements of squeezed and anti-squeezed quadratures, normalized to the vacuum noise level. We measured squeezing values of up to 4.8\,dB and anti-squeezing of up to 12.7\,dB. The sharp peaks in the spectrum originated from electronic pick up from the power supply due to antenna effects and a Pound-Drever-Hall (PDH) modulation frequency at 24\,MHz.

Using the squeezed light measurement we deduced the photo diode detection efficiencies to be around 99\,\% with a homodyne efficiency of  $\eta=\beta^{2}\cdot (1-\text{L})= 0.73$. This refers to a total detection efficiency of 0.725. The measured squeezing value between 5 and 100\,MHz is thus limited mainly by the homodyne visibility $\beta$ and the optical path losses L.

We again simulated the measured spectrum with the given parameters for the cavity (including linewidth), the pump field and the different homodyne detection efficiency (see section \ref{GHz-freq}) using N.L.C.S. The dashed lines in Fig. \ref{fig:Low-frequency-measurement} show the simulation for squeezing and anti-squeezing, respectively.

\section{Conclusion}
\begin{figure}[tb]
\centerline{\includegraphics[width=\textwidth]{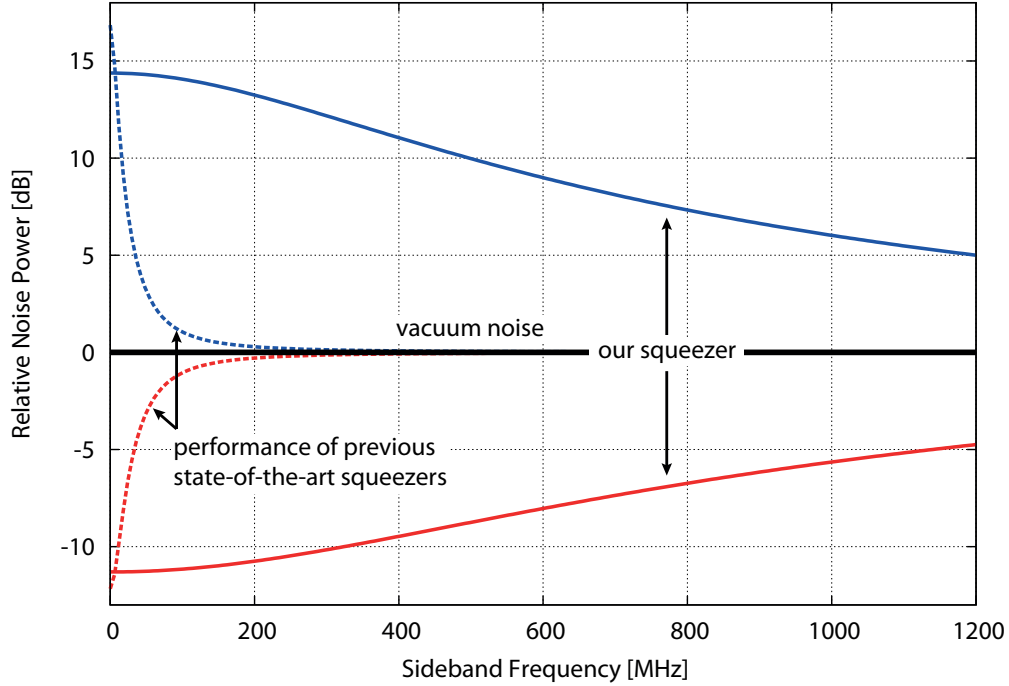}}
\caption{Numerical simulation using N.L.C.S. for a typical squeezing resonator as in \cite{Momo2012-12dB-1550} (dashed lines) and the monolithic GHz bandwidth squeezing resonator reported in this experiment (solid lines). Our simulation assumes a total detection efficiency of 96\,\%, as realized in \cite{Momo2012-12dB-1550} for low bandwidths. The squeezing source reported here generates almost the same squeezing strengths as state of the art sources. Its bandwidth does, however, offer significantly increased data rates in entanglement-based continuous-variable quantum key distribution.}
\label{fig:Hem-vs-Mono-Paper}
\end{figure}
Our work introduces a new, only 2.6\,mm long, monolithic crystal cavity. We demonstrated the generation of a broadband squeezed state at 1550\,nm ranging from 5\,MHz to 1.2\,GHz sideband frequency. Two different homodyne detectors were used to perform consecutive measurements for different frequency bands. We used the same optical parametric pump power of about 375\,mW for both measurements. A commercially available balanced photo receiver directly observed squeezing of up to 2\,dB between 10\,MHz and 1.2\,GHz. A second measurement used a home-made homodyne detector based on photo diodes with quantum efficiencies near unity. The latter directly observed a non-classical  quantum noise suppression of up to 4.8\,dB from 5 to 100\,MHz.

The two measured homodyne detector spectra where numerically simulated using identical parameters, but with different quantum efficiencies for the detectors. The simulations are self-consistent and in very good agreement with the measurements.

Based on our analysis, the measured squeezing was limited by the homodyne detector visibility (90\,\%), propagation loss (10\,\%) and partly by the quantum efficiencies of the photo diodes. Our analysis suggests that the current squeezing resonator design should enable the observation of squeezing up to about 10\,dB with a bandwidth in the GHz regime. Such a measurement would require a high-speed homodyne detector with GHz bandwidth, with 99\,\% detection efficiency and an increased homodyne visibility. We simulated our squeezing resonator setup with a total detection efficiency of 96\,\% as achieved in \cite{Momo2012-12dB-1550} for low bandwidths. The simulation shows around 10\,dB of non-classical noise suppression at MHz frequencies and more than 5\,dB at GHz frequencies (Fig.~\ref{fig:Hem-vs-Mono-Paper}).

Our squeezed-light resonator is a possible source for high-speed quantum key distribution. The source can be used to create two-mode squeezed states and therefore entanglement in the GHz band \cite{springerlink:10.1007/s11080-007-9030-x, 0295-5075-87-2-20005, QKD-Werner}. Our current design is capable of producing similarly high squeezing values as state of the art narrow-band squeezing resonators. The high squeezing bandwidth does, however, offer significantly improved QKD data rates. The squeezing strength is already limited by losses in fiber-based networks. Therefore, our scheme proposes a possible solution for high-speed quantum key distribution via squeezed states of light using optical fibers.

\section*{Acknowledgments}
The authors like to thank A. Samblowski, S. Steinlechner, V. H\"andchen, T. Eberle, H. Vahlbruch and A. R\"udiger. We acknowledge support from the International Max Planck Research School on Gravitational Wave Astronomy and the EU FP 7 project Q-ESSENCE (grant agreement no. 248095).


\begin{thebibliography}{10}
\newcommand{\enquote}[1]{``#1''}


\bibitem{PhysRevA.61.010303}
T.~C. Ralph, {\lq\lq}Continuous variable quantum cryptography," Phys. Rev. A \textbf{61}, 010303 (1999).

\bibitem{Madsen2012}
L.~S. Madsen, V.~C. Usenko, M. Lassen, R.~Filip, and U. L.~Andersen, {\lq\lq}Continuous variable quantum key distribution with modulated entangled states," Nat. Com.
  \textbf{3}, 1083 (2012).

\bibitem{Hillery2000}
M.~Hillery, {\lq\lq}Quantum cryptography with squeezed states," Phys. Rev. A \textbf{61}, 022309 (2000).

\bibitem{Silberhorn2002}
C.~Silberhorn, N.~Korolkova, and G.~Leuchs, "Quantum key distribution with bright entangled beams," Phys. Rev. Lett.
  \textbf{88}, 167902 (2002).


\bibitem{Cerf2005}
N.J.~Cerf, J.~Clavareau, C.~Macchiavello, and J.~Roland, {\lq\lq}Quantum entanglement enhances the capacity of bosonic channels with memory," Phys. Rev. A \textbf{72}, 042330 (2005).



\bibitem{PhysRevA.76.042305}
J.~Lodewyck, M.~Bloch, R.~Garc\'{i}a-Patr\'{o}n, S.~Fossier, E.~Karpov,
  E.~Diamanti, T.~Debuisschert, N.~J. Cerf, R.~Tualle-Brouri, S.~W. McLaughlin,
  and P.~Grangier, {\lq\lq}Quantum key distribution over 25km with an all-fiber continuous-variable system," Phys. Rev. A \textbf{76}, 042305 (2007).

\bibitem{Momo2010}
M.~Mehmet, T.~Eberle, S.~Steinlechner, H.~Vahlbruch, and R.~Schnabel, {\lq\lq}Demonstration of a quantum-enhanced fiber Sagnac interferometer," Opt.  Lett. \textbf{35}, 1665 (2010).

\bibitem{Tobi2011}
T.~Eberle, V.~H\"andchen, J.~Duhme, T.~Franz, R.~F. Werner, and R. Schnabel
  \enquote{Gaussian entanglement for quantum key distribution from a single-mode squeezing source,} arXiv:1110.3977v1.

\bibitem{PhysRevLett.102.130501}
R.~Garc\'{i}a-Patr\'{o}n and N.~J. Cerf, {\lq\lq}Continuous-variable quantum key distribution protocols over noisy channels," Phys. Rev. Lett. \textbf{102}, 130501
  (2009).

\bibitem{Slu85}
R.~E. Slusher, L.~W. Hollberg, B.~Yurke, J.~C. Mertz, and J.~F. Valley, {\lq\lq}Observation of squeezed states generated by four-wave mixing in an optical cavity," Phys. Rev. Lett. \textbf{55}, 2409 (1985).

\bibitem{PhysRevA.29.408}
B.~Yurke, {\lq\lq}Use of cavities in squeezed-state generation," Phys. Rev. A \textbf{29}, 408 (1984).

\bibitem{PhysRevLett.57.2520}
L.-A. Wu, H.~J. Kimble, J.~L. Hall, and H.~Wu, {\lq\lq}Generation of squeezed states by parametric down conversion," Phys. Rev. Lett. \textbf{57},
  2520 (1986).


\bibitem{Momo2012-12dB-1550}
M.~Mehmet, S.~Ast, T.~Eberle, S.~Steinlechner, H.~Vahlbruch, and R.~Schnabel, {\lq\lq}Squeezed light at 1550 nm with a quantum noise reduction of 12.3 dB," Opt. Exp.
  \textbf{19}, 25763 (2011).

\bibitem{Ast-Hem-GHz-Sqz}
S.~Ast, A.~Samblowski, M.~Mehmet, S.~Steinlechner, T.~Eberle, and R.~Schnabel, {\lq\lq}Continuous-wave nonclassical light with gigahertz squeezing bandwidth," Opt. Lett. \textbf{37}, 2367
  (2012).

\bibitem{PhysRevA.81.013814}
M.~Mehmet, H.~Vahlbruch, N.~Lastzka, K.~Danzmann, and R.~Schnabel, {\lq\lq}Observation of squeezed states with strong photon-number oscillations," Phys. Rev. A
  \textbf{81}, 013814 (2010).

\bibitem{Senior2007}
R.~J.~Senior, G.~N.~Milford, J.~Janousek, A.~E.~Dunlop, K.~Wagner, H-A.~Bachor, T.~C.~Ralph, E.~H.~Huntington, and C.~C.~Harb, {\lq\lq}Observation of a comb of optical squeezing over many gigahertz of bandwidth," Opt. Exp. \textbf{15}, 5310 (2007).

\bibitem{J-Steinlechner-Abs-2013}
J.~Steinlechner, S.~Ast, C.~Kr\"{u}ger, A.~P.~Singh, T.~Eberle, V.~H\"{a}ndchen, and R.~Schnabel, {\lq\lq}Absorption measurements of periodically poled potassium titanyl phosphate (PPKTP) at 775 nm and 1550 nm," Sensors \textbf{13}, 565-573, (2013). 


\bibitem{McKenzie2004}
K.~McKenzie, N.~Grosse, W.~P.~Bowen, S.~E.~Whitcomb, M.~B.~Gray, D.~E.~McClelland, and P.~K.~Lam, {\lq\lq}Squeezing in the audio gravitational-wave detection band," Phys. Rev. Lett. \textbf{93}, 161105 (2004).

\bibitem{Breitenbach1995}
G.~Breitenbach, T.~M\"uller, S.~F.~Pereira, J.-Ph.~Poizat, S.~Schiller, and J.~Mlynek, {\lq\lq}Squeezed vacuum from a monolithic optical parametric oscillator," J. Opt. Soc. Am. B \textbf{12}, 2304 (1995).

\bibitem{springerlink:10.1007/s11080-007-9030-x}
C.~Rod\`{o}, O.~Romero-Isart, K.~Eckert, and A.~Sanpera, {\lq\lq}Efficiency in quantum key distribution protocols with entangled gaussian states," Open Systs. Inf. Dyn.
  \textbf{14}, 69 (2007).

\bibitem{0295-5075-87-2-20005}
X.~Su, W.~Wang, Y.~Wang, X.~Jia, C.~Xie, and K.~Peng, {\lq\lq}Continuous variable quantum key distribution based on optical entangled states without signal modulation," EPL \textbf{87}, 20005
  (2009).

\bibitem{QKD-Werner}
F.~Furrer, T.~Franz, M.~Berta, A.~Leverrier, V.~B.~Scholz, M.~Tomamichel, and R.~F. Werner, {\lq\lq}Continuous variable quantum key distribution: Finite-key analysis of composable security against coherent attacks," Phys. Rev. Lett. \textbf{109}, 100502 (2012).

\end{thebibliography}
\end{document}